\documentclass[sn-mathphys]{sn-jnl}
\usepackage{bm}
\usepackage{graphicx}
\usepackage{amsmath}
\usepackage{ucs}
\usepackage[utf8x]{inputenc}
\usepackage{textcomp}
\usepackage{siunitx}
\DeclareSIUnit\Molar{\textsc{m}}

\begin{document}

\title[ ]{Memory induced Magnus effect}

\author[1]{\fnm{Xin} \sur{Cao}}
\author[2]{\fnm{Debankur} \sur{Das}}
\author[1]{\fnm{Niklas} \sur{Windbacher}}
\author[1]{\fnm{Felix} \sur{Ginot}}
\author[2]{\fnm{Matthias} \sur{Kr\"uger}}
\author*[1]{\fnm{Clemens} \sur{Bechinger}}\email{clemens.bechinger@uni-konstanz.de}
\affil*[1]{Fachbereich Physik, University Konstanz, 78464 Konstanz, Germany}
\affil[2]{Institut für Theoretische Physik, Universität Göttingen, 37077 Göttingen Germany}

\abstract{Spinning objects which move through air or liquids experience a Magnus force. This effect is commonly exploited in ball sports but also of considerable importance for applications and fundamental science. Opposed to large objects where Magnus forces are strong, they are only weak at small scales and eventually vanish for overdamped micron-sized particles in simple liquids. Here we demonstrate an 
about one-million-fold enhanced Magnus force 
of spinning colloids in viscoelastic 
fluids. Such fluids are 
characterized by a time-delayed response to external perturbations which causes a deformation of the fluidic network around the moving particle. When the particle additionally spins, the deformation field becomes misaligned relative to the particle’s moving direction, leading to a force perpendicular to the direction of travel and the spinning axis. The presence of strongly enhanced memory-induced Magnus forces at microscales opens novel applications for particle sorting and steering, the creation and visualization of anomalous flows and more.}


\maketitle

When a spinning object travels through a fluid, its trajectory is typically curved. Although Isaac Newton was the first to describe this effect in 1671 \cite{newton1993new}, it is commonly attributed to Heinrich Gustav Magnus who provided a physical explanation on the influence of rotation on the motion of objects \cite{magnus1853ueber}. Nowadays the Magnus effect is well established and finds use not only in ball games but is also exploited, e.g., as economic propulsion mechanism for ships \cite{de2016flettner,bordogna2019experiments,bordogna2020effects,seddiek2021harnessing} or to provide lifting forces for air vehicles \cite{seifert2012review}. In addition to applications, Magnus effects are also relevant for the understanding of planet formation inside proto-planetary discs \cite{forbes2015curveballs}, the behavior of ions in superfluids \cite{donnelly1969stochastic,sonin1997magnus} and are even discussed in context of the motion of vortex lines in superconductors\cite{ao1993berry}. In general, the Magnus force results from an asymmetry of the velocity field in the medium around a translating and simultaneously rotating object. According to the Bernoulli equation this results in pressure inhomogeneities near the object and a force perpendicular to the direction of travel, i.e., the Magnus force $\mathbf{F}_\mathrm{M}=f (\bm{\omega}\times\mathbf{v})$. 
Here, the Magnus coefficient $f$ quantifies the coupling of the particle to its surrounding and
$\bm{\omega}$ and $\mathbf{v}$ are the angular and the linear velocities of the object relative to the fluid, respectively. While in most cases $f>0$, it can be also negative, e.g. at high velocities where the flow around the object is partially turbulent \cite{seifert2012review,kim2014inverse}, or when spinning objects move through rarefied gases or granular media \cite{borg2003force,kumar2019magnus,seguin2022forces}. 
Opposed to large objects where Magnus forces can be very strong, they are weak at small scales. In case of Brownian, i.e. micron-sized particles in simple fluids, they eventually vanish since viscous forces dominate over inertial effects \cite{changfu2003lift,solsona2020trajectory}. Therefore, applications of Magnus forces in such systems are rare.

Here, we report the experimental observation of a strong memory induced Magnus effect for spinning micron-sized colloidal particles moving through an overdamped viscoelastic fluid. Opposed to viscous, i.e. Newtonian liquids, which instantaneously respond to external perturbations, viscoelastic fluids are characterized by stress-relaxation times $\tau$ on the order of seconds and beyond \cite{dhont1996introduction,larson1999structure}. 
Similar to the Magnus force $\mathbf{F}_\textrm{M}$ generated in a viscous liquid, in viscoelastic liquid the memory-induced Magnus force $\mathbf{F}_\mathrm{mM}=\tilde{f} (\bm{\omega}\times\mathbf{v})$ is exerted on a translating and spinning object. The coefficient $\tilde{f}<0$ and its amplitude is larger than $f$ (i.e. the coefficient in a pure viscous fluid) by a factor more than $10^6$.
Our experimental results are in excellent agreement with a theoretical description, where the time-delayed response of the fluid around a moving object is modelled by a density dipole. While this dipole points in the direction of $\mathbf{v}$ for a pure translational particle motion, it is rotated when the particle exhibits an additional spinning motion. As a result, a force component perpendicular to the driving force arises which eventually leads to $\mathbf{F}_\mathrm{mM}$. Such model also explains our experimental finding that, when the particle's spinning motion is stopped, the force $\mathbf{F}_\mathrm{mM}$ remains present, and only decays after time  $\tau$. Since our findings should apply to a large number of viscoelastic fluids, we expect that this unusual type of Magnus force leads to novel applications, e.g. in the field  of particle sorting and steering but also the creation and visualization of anomalous flows~\cite{banerjee2017odd,souslov2019topological,yang2021topologically,kalz2022collisions,reichhardt2022active}.

\begin{figure*}[ht!]
  \centering
  \includegraphics[width=1.0\columnwidth]{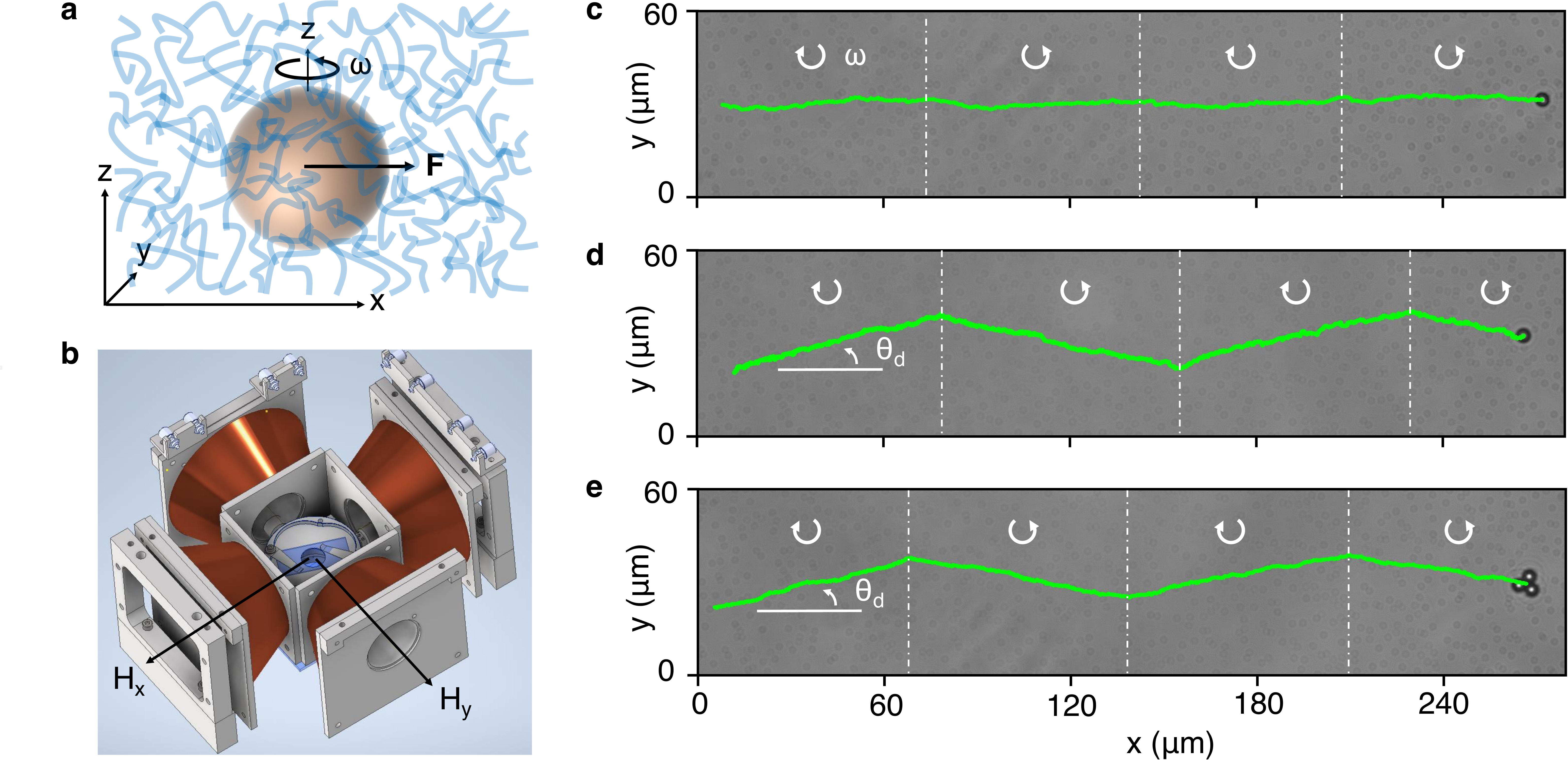}
\caption{\textbf{Memory induced Magnus force acting on colloidal particles in viscous and viscoelastic fluids} \textbf{a,} Illustration of a super paramagnetic colloid sphere spinning at angular frequency $\omega$ (rotation axis in $z$ direction) and driven by a force $\mathbf{F}$ along $x$ direction through a viscoelastic fluid. \textbf{b,} Sketch of the experimental setup, with two perpendicular pairs of magnetic coils generating a rotating magnetic field $\mathbf{H}(t)$ with frequency $\omega_\mathrm{H}$, in the $xy$ plane. Due to a magnetic torque the colloid is set into a spinning motion with $\omega \ll \omega _\mathrm{H}$. 
\textbf{c,d,} Trajectories (green) of spinning colloidal particles driven through a fluid with $H=732$ A/m, $\omega _\mathrm{H}=20\pi~\mathrm{s}^{-1}$ and $F=156$ fN. The sign of the spinning direction is periodically reversed as indicated by the curved arrows. \textbf{c,} Single colloidal particle in a purely viscous water-glycerol (1:1 by weight) mixture for a colloid driving velocity $v_x=0.288\mu\mathrm{m/s}$. \textbf{d,} Single colloidal particle in a \SI{5}{\milli\Molar} micellar solution with $v_x=0.065\mu\mathrm{m/s}$ and $\theta_\mathrm{d}\approx15.0^{\circ}$.  \textbf{e,} Colloidal trimer in the same micellar solution with $v_x=0.178\mu\mathrm{m/s}$ and $\theta_\mathrm{d}\approx14.4^{\circ}$.
}
\label{fig1}
\end{figure*}

In our experiments we are using super paramagnetic colloidal spheres (diameter $\sigma = 4.45~\mu\mathrm{m}$) which are suspended in a viscoelastic fluid (see below) and contained within a thin sample cell. 
The motion of particles is imaged with a video camera mounted on an inverted microscope and analyzed using digital video microscopy. 
The microscope is mounted on a tilting stage which allows to exert on each particle an external (gravitational) drift force $\mathbf{F}=mg\sin\alpha~\hat{\mathbf{x}}$. 
Here $mg=286~\mathrm{fN}$ is the buoyant particle weight, $\alpha$ the tilting angle and $\hat{x}$ the unit vector along $x$ direction. 
The angle $\alpha$ can be tuned between $0^{\circ}$ and $35^{\circ}$. 
Additionally, we use two perpendicular pairs of coils (Figs. \ref{fig1}a,b) which create a rotating magnetic field $\mathbf{H}(t)$ in the sample plane with components $H_x(t) = H \cos\omega_\mathrm{H}t$ and $H_y(t) = H \sin\omega_\mathrm{H}t$.
The frequency $\omega_\mathrm{H}=20\pi~\mathrm{s}^{-1}$ is fixed in our experiments. 
The rotating $\mathbf{H}$ induces a rotating magnetization $\mathbf{M}$ within the colloidal particles. 
Due to a phase lag in $\mathbf{M}$, the rotating magnetic field applies a torque $\Gamma = \vert \mathbf{M}\times\mathbf{H}\vert \propto H^2$ to the colloid spheres.
This results in an additional spinning motion with angular frequency $\omega\propto\Gamma$ (see Methods and Supplementary Fig. 1 for details). 

Two different types of viscoelastic fluids were used in our experiments: (i) an entangled giant worm-like micellar solution \cite{cates1990statics} composed of about $5$ mM per liter equimolar cetylpyridinium chloride monohydrate (CPyCl) and sodium salicylate (NaSal) dissolved in water and (ii) a semi-dilute aqueous polymer solution of polyacrylamide (PAAM) with molecular weight \SI{18}{\mega\dalton} and mass concentration 0.03\%. 
Both fluids exhibit a pronounced viscoelastic behavior as confirmed by microrheological experiments \cite{narinder2018memory,ginot2022recoil} (see Methods and Supplementary Fig. 2 for details). 
All our experiments have been performed at a constant sample temperature of \SI{25}{\celsius}. 

To demonstrate that conventional Magnus forces in colloidal systems are vanishingly small when suspended in merely viscous, i.e., Newtonian fluids \cite{solsona2020trajectory, changfu2003lift}, we first studied the motion of a colloidal particle in a water-glycerol mixture. 
Figure~\ref{fig1}c shows the corresponding trajectory of a particle subjected to a rotating magnetic field (H=732 A/m) and a driving force $F=156$ fN. 
The rotation direction of $\mathbf{H}$ has been periodically reversed from clockwise to anti-clockwise direction to rule out the possible influence of a particle drift in y-direction. Within our experimental resolution, no particle deflection from the direction of the driving force is observed. This is consistent with the theoretically predicted conventional Magnus force in viscous liquids in the limit of low Reynolds number $\mathbf{F}_\mathrm{M} = \pi\rho(\sigma/2)^3\bm{\omega}\times\mathbf{v}$, where $\rho$ is the fluid mass density \cite{rubinow1961transverse,solsona2020trajectory}. For our liquid this yields a deflection angle $\theta_\mathrm{d}=\arctan(F_\textrm{M}/F)\ll0.0002^{\circ}$ (considering $\omega\ll\omega_\mathrm{H}$), which is below the experimental resolution. When repeating the experiment in a viscoelastic fluid, however, a pronounced deflection of the trajectory is observed (Fig.~\ref{fig1}d). 
From the measured velocity ratio $v_y/v_x$ we determine the deviation angle $\theta_\mathrm{d}=\arctan(v_y/v_x)\approx15^{\circ}$. The deflection changes its direction upon reversing the direction of rotation of $\mathbf{H}$. 
Although we are particularly interested in the angular motion of the spinning particles, it can not be resolved for the case of single colloidal spheres.
Therefore, in the following we use colloidal trimers whose angular motion is easily measured, while showing almost identical behavior (Fig.~\ref{fig1}e and Supplementary Videos 1 and 2). 
For details regarding the formation of such trimers we refer to the Methods section.
For trimers with a spinning frequency $\omega=0.97$ rad/s and drift velocity $v_x=0.16~\mu\mathrm{m/s}$, we find an angular deflection $\theta_\mathrm{d}\approx14.4^{\circ}$ which demonstrates a force pointing in the direction $-\bm{\omega}\times\mathbf{v}$.

\begin{figure*}[t]
  \centering
  \includegraphics[width = \columnwidth]{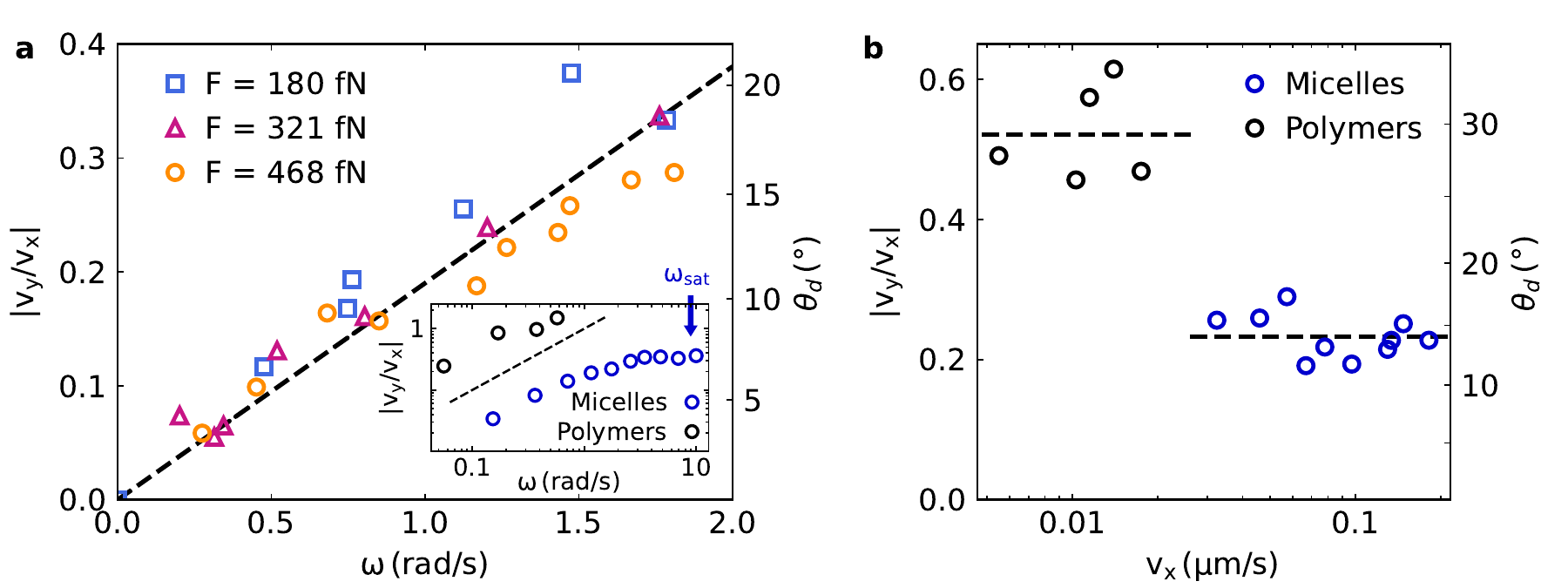}
\caption{{\textbf{Dependence of translational and angular velocities.}} \textbf{a,} Measured velocity ratio $\vert v_y/v_x\vert $ (left axis) and deflection angle $\vert \theta_\mathrm{d}\vert $ (right axis) of a colloidal trimer in a micellar fluid as a function of its rotational speed $\omega$ for an applied force $F=468~\mathrm{fN}$ (orange circles), $321~\mathrm{fN}$ (purple triangles) and $180~\mathrm{fN}$ (blue squares), respectively. The dashed line is a fit to all data with equation $\vert v_y/v_x\vert =k\omega$ and the fitting parameter $k=0.201\pm0.008$ s. Inset: the $\vert v_y/v_x\vert $ as a function of $\omega$ for a colloidal trimer driving through viscoelastic PAAM (black circles) and micellar solutions (blue circles). The dashed line indicates a linear relation between $\vert v_y/v_x\vert $ and $\omega$. The vertical arrow indicates, for micellar solution, the theoretically predicted angular frequency where $\vert v_y/v_x\vert $ saturates. \textbf{b,} $\vert v_y/v_x\vert $ and $\vert \theta_\mathrm{d}\vert $ versus the drift velocity $v_x$ for colloidal trimer in PAAM (black circles) and micellar solutions (blue circles). The data in PAAM and micellar solutions were obtained at magnetic fields $H=976$ A/m and $H=732$ A/m, respectively, which corresponds to a rotational trimer velocity of $\omega\approx0.17$ rad/s and $\omega\approx1.2$ rad/s. Dashed lines indicate $\vert v_y/v_x\vert =0.52$ and $\vert v_y/v_x\vert =0.23$. 
}
\label{fig2}
\end{figure*}

To quantify our observations, we measured the velocity ratio $v_y/v_x$ as a function of the trimer spinning velocity $\omega$, and for different driving forces $\mathbf{F}$ (Fig.~\ref{fig2}a). 
At small $\omega$ we observe for both viscoelastic fluids a linear behavior which eventually saturates for larger $\omega$ (see  Fig.~\ref{fig2}a, inset). In addition, $v_y/v_x$ is independent of $v_x$ when $\omega$ is kept constant (Fig.~\ref{fig2}b). 
Both findings suggests that in the linear regime a memory-induced Magnus force $\mathbf{F}_\mathrm{mM}=\tilde{f} ( \bm{\omega}\times\mathbf{v})$ is acting on the colloidal trimer with prefactor $\tilde{f}$ depending on the colloid-fluid interaction.

\begin{figure*}[!ht]
  \centering
  \includegraphics[width=\textwidth]{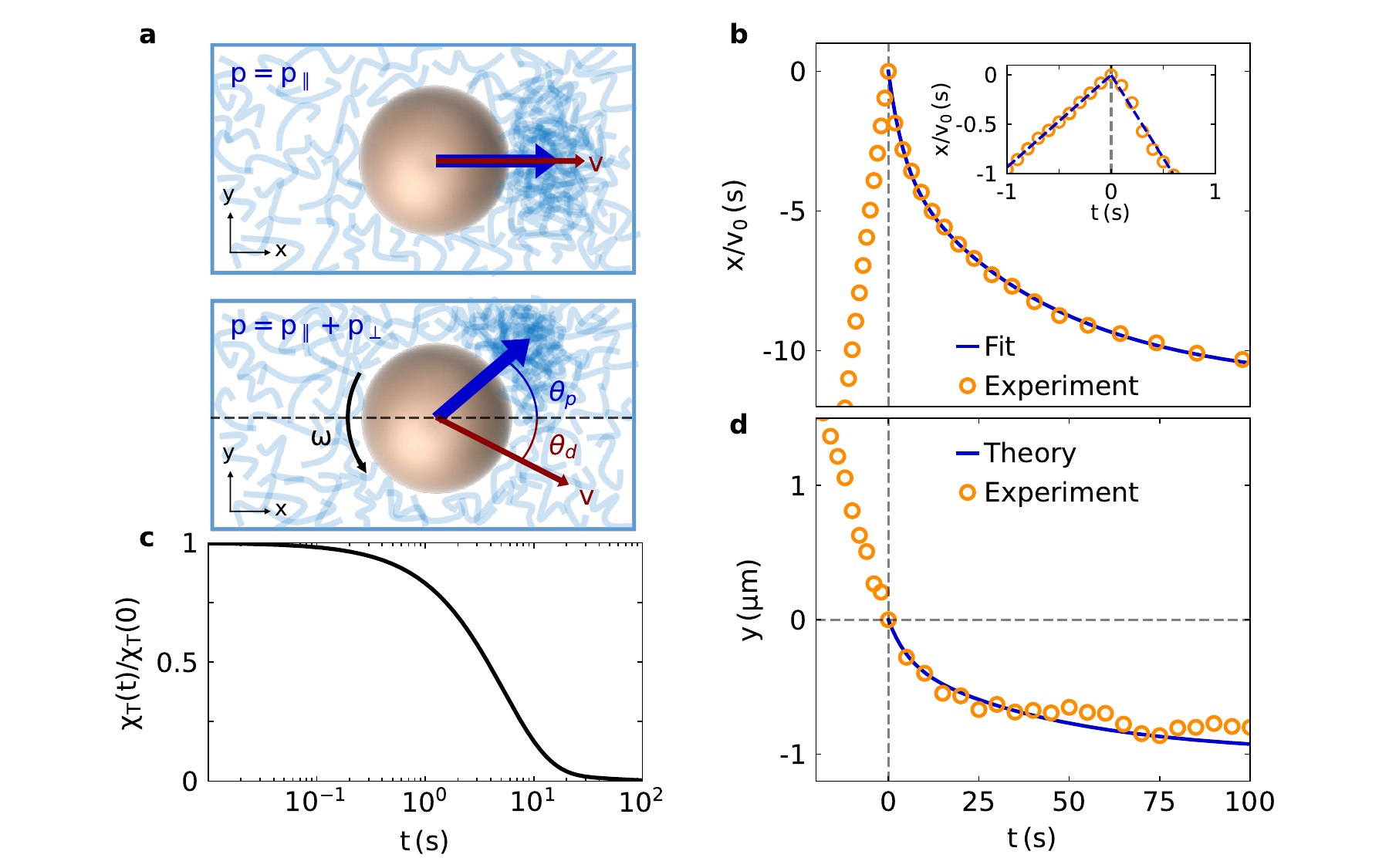}
\caption{\textbf{Theoretical model}. \textbf{a,} Top: Illustration of a colloidal particle moving at linear velocity $v$ in a viscoelastic fluid. The linear motion deforms the microscopic network structure of the liquid, leading to a density gradient as characterised by the density dipole $\mathbf{p}_{\parallel}$ parallel to $\mathbf{v}$. Bottom: When the particle is additionally set into rotation $\bm{\omega}$, the dipole is rotated within the $xy$-plane by an angle $\theta_\mathrm{p}$, which leads to the deflection of the particle trajectory by an angle $\theta_\mathrm{d}$. \textbf{b,} Measured $x$ displacement of a non-spinning trimer (orange symbols) during a recoil experiment where the trimer is first driven with velocity $v_0 = 0.95~\mu\mathrm{m/s}$ through the micellar fluid and the applied force is suddenly removed at $t,x=0$. The data are normalised by $v_0$. The solid line represents a fitting to the equation $x=a_1\exp(-t/\tau_1)+a_2\exp(-t/\tau_2)-a_1-a_2$, with $a_1=3.14~\mu\mathrm{m}$, $\tau_1=5.20$ s, $a_2=7.60~\mu\mathrm{m}$, $\tau_2=44.5$ s.
Inset: magnified region near $t=0$. From the ratio of the slopes (as indicated by the dashed lines) after and before $t=0$ we obtain the recoil ratio $r=1.8\pm0.1$. \textbf{c,} The memory kernel $\chi_T(t)$ (normalised by $\chi_T(0)$) as defined in Eq.~(\ref{eq_trans_recoil}), which is derived from the fitted curve shown in (\textbf{b}). \textbf{d,} Measured (orange symbols) $y(t)$ displacement of a trimer in a micellar fluid which is subjected for $t<0$ to a driving force $F=468$ fN and a torque $\Gamma = 37.3\mathrm{pN}\cdot\mu\mathrm{m}$. After setting $\Gamma=0$ at $t=0$ the displacement in $y$ direction only decays after about 44.5 s. The data are averaged over 18 measurements. The solid line is the theoretically predicted trajectory $y(t)=0.093[a_1\exp(-t/\tau_1)+a_2\exp(-t/\tau_2)-a_1-a_2]$ with $a_1$, $\tau_1$, $a_2$, $\tau_2$ identical as in (\textbf{b}). 
}
\label{fig3}
\end{figure*}

For a theoretical understanding of the above observations, we first consider a non-spinning particle moving at velocity $\mathbf{v}$ due to an external force $\mathbf{F}$ through a viscoelastic fluid. Owing to the finite stress-relaxation time $\tau$, a fore-rear inhomogeneity within the fluid builds up around the particle. It can be characterized by a density dipole $\mathbf{p}_{\parallel}$ pointing in the direction of $\mathbf{v}$ (Fig.~\ref{fig3}a, top and Supplementary Note 1)~\cite{dhont1996introduction,dzubiella2003depletion,squires2005simple,rauscher2007dynamic,khan2019optical}. Within linear response theory ~\cite{chaikin1995principles}, the time-dependent magnitude of $\mathbf{p}_{\parallel}$ is given by a history integral of the force, 

\begin{eqnarray}
p_{\parallel}(t) = \int_{-\infty}^{t}\chi_T(t - t^\prime) F(t') dt',
\label{eq_trans_recoil}
\end{eqnarray} 
where $\chi_T(t)$ is a memory kernel that characterizes the dipole relaxation dynamics~\cite{kubo1957statistical}. Experimentally, $\chi_T(t)$ can be determined 
when the driving force acting on the particle is suddenly removed. This results in a restoring force anti-parallel to $\mathbf{p}_{\parallel}$ leading to a recoil motion opposite to the driving direction. Fig.~\ref{fig3}b (also Supplementary Videos 3 and 4) shows such recoil for a colloidal trimer driven by a magnetic field gradient force that was turned off at $t=0$. In agreement with previous studies, the recoil is well described by a double-exponential decay \cite{gomez2015transient,khan2019optical}. From such data we immediately obtain the memory kernel $\chi_T(t)$ as shown in Fig.~\ref{fig3}c (see details in Supplementary Note 2).

When the particle is additionally set into a spinning motion by an external torque $\Gamma$, the orientation of the density dipole in the $xy$-plane changes by an angle $\theta_\mathrm{p}$ (Fig.~\ref{fig3}a bottom) due to the particle-fluid interaction. Accordingly, one obtains a dipole 
$\mathbf{p}=\mathbf{p}_{\parallel}+\mathbf{p}_\perp$ with $\mathbf{p}_\perp$ the component perpendicular to $\mathbf{F}$ . Similar to the restoring force which is caused by $\mathbf{p}_{\parallel}$, a force perpendicular to $\mathbf{F}$ is caused by $\mathbf{p}_\perp$. This perpendicular force is identified as the memory-induced Magnus force $\mathbf{F}_\mathrm{mM}$ which is proportion to $-\mathbf{p}_\perp$. 
We consider the regime where $\mathbf{p}_\perp$ is a linear function of $\mathrm{F}$ and $\Gamma$, for which we find (see details in Supplementary Note 1)
\begin{eqnarray}
p_{\perp}(t) &=&  \int_{-\infty}^{t} dt^\prime \chi_R(t - t^\prime) \Gamma(t') p_{\parallel}(t').
    \label{eq_orient_recoil}
    \end{eqnarray}
Here $\chi_R(t)$ is the memory kernel associated with the relaxation of the dipole component $p_{\perp}$. 
Because in the range of our experiments we have $\Gamma\propto\omega$ and $p_{\parallel}\propto F\propto v_x$ (see Supplementary Fig. 3), this leads to $p_\perp \propto \omega v_x$ or $\mathbf{F}_\mathrm{mM}=\tilde{f} (\bm{\omega}\times\mathbf{v})$, in agreement with the observations in Fig.~\ref{fig2}. 
As a side note we mention that after the particle starts to move in $y$ direction under the influence of $\mathbf{F}_\mathrm{mM}$, this creates - similar to $\textbf{p}_{\parallel}$ - an additional density dipole which is opposite to $\textbf{p}_{\perp}$ and reduces the velocity component $v_y$. This effect is already included in Eq.~(\ref{eq_orient_recoil}) by considering $\chi_R(t)$ for a moving particle.

According to Eq.~(\ref{eq_orient_recoil}), $p_{\perp}$ (and hence $\mathbf{F}_\mathrm{mM}$) should not instantaneously vanish when removing the external torque applied to the particle. 
This means, the particle motion in $y$ direction decays on a timescale given by $\chi_R(t)$ which characterizes how fast $\mathbf{p}_{\perp}$ decays to zero. To demonstrate this, Fig.~\ref{fig3}d shows for $t<0$ the motion of a trimer in $y$ direction under the influence of a force $F=468$ fN and torque $\Gamma = 37.3\mathrm{pN}\cdot\mu\mathrm{m}$. When $\Gamma$ is set to zero at $t=0$ s, the cluster's motion in $y$ direction decays only on a timescale of several ten seconds. This decay can be directly compared to the corresponding prediction (solid line) of Eq.~(\ref{eq_orient_recoil}) when assuming $\chi_R(t) \propto \chi_T(t)$. Such proportionality is plausible, because the susceptibilities that describe relaxations of dipole fields in different directions can be expected to be similar due to isotropy of the quiescent fluid. 

Based on the dipole-rotation picture as described in Fig.~\ref{fig3}a, we also construct a simple Maxwell-like model for the time dependence of the dipole orientation $\phi$, which contains a driving term proportional to the applied torque $\Gamma$ and a restoring term due to the relaxation of $\mathbf{p}_{\perp}$, i.e.
\begin{eqnarray}
\dot{\phi} &=&  C\frac{\Gamma}{\gamma_R}-\frac{\phi}{\tau}
\label{eq_orient_dipole2}
\end{eqnarray}
    with $\gamma_R$ the steady state rotational friction coefficient, so that $\Gamma/\gamma_R=\omega$ is the rotational velocity of the colloid, and  $C\leq 1$ a constant which describes the coupling between the rotation of the colloid and that of the density dipole. In absence of $\Gamma$, the solution of  Eq.~(\ref{eq_orient_dipole2}) relaxes exponentially with time scale $\tau$. Under steady state driving conditions the value of the dipole orientation angle $\phi$ becomes
\begin{eqnarray}
\theta_\mathrm{p} =C\omega\tau.
\label{eq_orient_dipole}
\end{eqnarray}
Considering that the dipole force, the viscous force and the external force must balance altogether, this leads to a simple relationship between $\theta_\mathrm{p}$ and the velocity ratio $v_y/v_x$ (see Supplementary Note 3 for details):

\begin{eqnarray}
    \left\vert \frac{v_y}{v_x}\right\vert =r \theta_\mathrm{p}= r C\omega \tau.\label{eq:v}
\end{eqnarray} 
Here, $r$ is the dimensionless {\it recoil ratio} given by the ratio of the particle velocities before and right after turning off the driving force. From the typical recoil shown in Fig.~\ref{fig3}b for the micellar system we determine $r\approx 1.8\pm0.1$. A similar analysis can be also applied to the PMMA solutions to obtain the corresponding $r$ (see Supplementary Fig. 2). 
Eq.~(\ref{eq:v}) also allows to estimate $\tilde f$ defined above by interpreting the motion in $y$ to arise from the force ${\bf F}_{\rm mM}$. It yields $\tilde f=-Cr\tau\gamma$, with $\gamma=6.50\pm0.09~\mathrm{pN}\cdot\mathrm{s}/\mu\mathrm{m}$ the translational friction coefficient (see Supplementary Fig. 3). By fitting Eq.~(\ref{eq:v}) to the linear part of our data shown in Fig.~\ref{fig2}a, we obtain $rC\tau=0.201\pm0.008$ s, or $\tilde f=1.29\pm0.06~\mathrm{pN}\cdot\mathrm{s}^2/\mu\mathrm{m}$. 
This is considerably larger than the Magnus coefficient for a viscous liquid, which is $f=\pi\rho(2\sigma/2)^3=0.28\times10^{-6}~\mathrm{pN}\cdot\mathrm{s}^2/\mu\mathrm{m}$ (for this estimate we have approximated the cluster of three colloids, each with diameter $\sigma$, with a single spherical particle with diameter $2\sigma$).
Within the above framework we also obtain a simple estimate for the upper limit of $\omega$ where the linear relation between $\vert v_y/v_x\vert $ and $\omega$ ends. Assuming that the magnitude of $p_\perp$ cannot exceed that of the original dipole, this yields  $\vert v_y/v_x\vert < r$ (see Supplementary Note 3). From this one obtains a rough estimate of the saturation velocity $\omega_{\rm sat}\leq 1/C\tau\approx9.0~ \mathrm{s}^{-1}$ which is consistent with our data in the inset of Fig.~\ref{fig2}a. Following a similar procedure we obtain, for a colloidal trimer in a polymer fluid, a saturation velocity about $\omega_{\rm sat}\leq 1/C\tau\approx3.4~ \mathrm{s}^{-1}$.

As an alternative approach to Magnus forces one can introduce a viscosity tensor ${\boldsymbol{ \eta}}_{\mathrm{mM}}$ which is defined by ${\bf F}_{\mathrm{mM}}=  \tilde f (\bm{\omega}\times\mathbf{v})=-3\pi \sigma {\boldsymbol{ \eta}_{\mathrm{mM}}}\cdot {\bf v}$. Writing the cross product using the Levi-Civita symbol, $F_{\mathrm{mM},i}=-\tilde f \omega e_{3ij} v_j$,  this shows ${\boldsymbol{ \eta}}_{\mathrm{mM}}$ is anti-symmetric. When additionally considering the translational friction force $-\gamma {\bf v}$ to obtain the diagonal components of the viscosity tensor
  with the friction coefficient $\gamma$ for $\omega=0$, introduced below Eq.~(\ref{eq:v}), this yields, for ${\bf \omega}$ pointing in $z$-direction,
\begin{eqnarray}
  3\pi \sigma  \eta_{ij}=
  \begin{pmatrix}
         \gamma&  \tilde f\omega  \\
         -\tilde f\omega& \gamma
    \end{pmatrix}.
\end{eqnarray}

Such non-symmetric viscosity tensors (typically referred to as odd viscosity ~\cite{banerjee2017odd,souslov2019topological,yang2021topologically,kalz2022collisions, reichhardt2022active,markovich2021odd, lou2022odd}) which generally results from the violation of Onsager's reciprocal relations~\cite{onsager1931reciprocal1,onsager1931reciprocal2} is in our system caused by the time-delayed dynamics of the density dipole. Accordingly, the off-diagonal elements of the viscosity tensor can be easily tuned via the rotation frequency.

In summary, our work demonstrates that Magnus forces which typically vanish in the realm of small Reynolds number can be very strong in case of viscoelastic fluids which exhibit a time-delayed response to perturbations. In addition to rotating magnetic fields which have been used in our study to induce a spinning motion, also electrical \cite{fennimore2003rotational} or optical fields \cite{kuhn2017optically} have been demonstrated to impose considerable torques on particles down to the nanometer scale. This allows to extend the use of Magnus forces to the regime of small Reynold numbers which may lead to new types of microswimmers, novel strategies for steering and sorting of particles but also to visualize complex flow patterns in liquids.

\section*{Methods}
\subsection*{Sample preparation.} We prepare highly diluted colloidal suspensions of superparamagnetic spheres (Dynabeads M-450, diameter $\sim\SI{4.5}{\micro\m}$) dispersed in a viscoelastic fluid. The number density (roughly $10^8$ per liter) corresponds to less than ten colloid spheres within our field of view $ \SI{240}{\micro\metre} \times \SI{300} {\micro\metre} $. For viscoelastic fluid we use either a polymer solution or a micellar solution. The polymer solution is a semi-dilute aqueous solution of poly-acrylamide (PAAM) with molecular weight \SI{18}{\mega\dalton} and mass concentration 0.03\%. The micellar solution is an equimolar, aqueous solution of cetylpyridinium chloride monohydrate (CPyCl) and sodium salicylate (NaSal). The molar density is \SI{5.5}{\milli\Molar} per liter for data in Fig.2a and \SI{5.0}{\milli\Molar} per liter for Fig.2a inset and all other relevant data.
To make a colloid sample, we inject the colloid suspension into a glass sample cell of about $\SI{20}{\milli\metre} \times \SI{10} {\milli\metre}\times \SI{0.2} {\milli\metre} $ in size, where \SI{0.2}{\milli\m} is the sample thickness. After the sample is made, it is then transferred to an inverted Nikon microscope where we can observe the motion of the colloid. During the experiment the temperature of the sample is kept at $25 \pm \SI{1}{\celsius}$ using a flow thermostat.

\subsection*{Formation of colloidal trimers.} The rotating magnetic field creates an effective long range attraction ($\propto r^3$) between the magnetic colloid \cite{martinez2015magnetic}. In presence of a rotating magnetic field with $H=\SI{732}{\A\per\m}$, colloidal particles with distances of $\sim 5\sigma$ experience a strong attraction and form dense clusters, see Supplementary Video 5. To increase the chance of particle encounters during the cluster formation process, the sample stage has been tilted by $\sim$10 degrees. This leads to the formation of rigid and stable colloidal clusters with random sizes including trimers. 

\subsection*{Calibration of magnetic torques.} The spinning of the colloid and its aggregates are induced via a magnetic torque $\Gamma=\gamma_\mathrm{m}H^2$ applied by the rotating magnetic field as shown in Fig. 1b. 
In steady rotation, the magnetic torque balances the viscous torque of the fluid, i.e. $\Gamma=\gamma_\mathrm{R}\omega$, where $\gamma_\mathrm{R}=6\pi\eta \sigma^3$ for a colloid trimers \cite{cao2022moire}. This leads to $\omega=kH^2$, where $k=\gamma_\mathrm{m}/\gamma_\mathrm{R}$.
To calibrate the magnetic torque. We first rotate the colloid trimer in water, where the measured rotating speed $\omega$ of a colloid trimer as a function of $H$ follows exactly the equation $\omega=kH^2$ as shown in Supplementary Fig. 1, with the fitted $k=3.26\times10^{-5}~\mathrm{m}^2/(\mathrm{A}^2\cdot\mathrm{s})$.  
Considering the measured $\eta=1.29\times10^{-3}~\mathrm{Pa}\cdot\mathrm{s}$ for water and the colloid diameter $\sigma=4.45~\mu\mathrm{m/s}$, we obtain $\gamma_\mathrm{m}=6.99\times10^{-5}~\mathrm{pN}\cdot\mu\mathrm{m}/(\mathrm{A/m})^2$.  

\subsection*{Colloidal recoils with time-dependent magnetic gradients.} Because gravitational drift forces on the particles can not be suddenly changed, the translational recoil curves as shown in Fig. 3b have been measured by using a permanent magnet whose position within the sample plane could be suddenly changed ($<$0.1s) within the sample plane with a mechanical spring-loaded device. To do so we first put the magnet close to the sample cell where the magnetic field gradient leads to a constant drift velocity of the trimers. Upon activating the spring-loaded retraction mechanism, the magnet is pulled away and the colloidal recoil sets in.

\section*{Acknowledgement}
We acknowledge helpful discussions with Matthias Fuchs, Pietro Tierno and Gaspard Junot. This work is funded by the Deutsche Forschungsgemeinschaft (DFG), Grant No. SFB 1432 - Project ID 4252172. F.G. acknowledges support from the Humboldt foundation.

\makeatletter
\setcounter{equation}{0}

\section*{Author contributions}
C.B. and X.C. designed the experiments which were carried out and analyzed by X.C. and N.W. The theoretical model has been developed by D.D. and M.K.. F.G. contributed to the overall discussions. All authors contributed to the writing of the paper.

\section*{Competing Interests}
The authors declare no competing interests

\section*{Data Availability}
Raw data of this work is available from the corresponding author on reasonable request.

\bibliography{main}

\clearpage




\section{Supplementary Information for ``Memory induced Magnus effect"}

\maketitle

\makeatletter
\renewcommand{\theequation}{S\@arabic\c@equation}
\makeatother

\subsection*{Supplementary Note 1: Magnus velocity and density dipoles}
The micellar concentration adjacent to a single colloidal particle or cluster can be expanded as  
\begin{equation*}
    \rho(\mathbf{r}) = \sum_{l = 0}^{\infty} \sum_{m=-l}^{l} Y_{lm}(\theta,\phi) R_{lm}(r) C_{lm}.
\end{equation*}
Here $\mathbf{r} = \{r,\theta,\phi\}$ denotes the distance (in spherical coordinates) away from the colloid center, and we have expanded $\rho$ in normalized radial functions $R_{lm}(r)$ and (normalized) angular parts $Y_{lm}(\theta,\phi)$  (the real spherical harmonics). The relevant spherical harmonics for the Magnus effect are $Y_{11}(\theta, \phi)$ and $Y_{1,-1}(\theta,\phi)$ which correspond to dipoles along $x$ and $y$ directions respectively. We thus identify the coefficients corresponding to $Y_{11}(\theta, \phi)$ and $Y_{1,-1}(\theta,\phi)$ as $p_{\parallel} = C_{11} $ and $p_{\perp} = C_{1-1}$ respectively.

The spherical harmonics $Y_{11}(\theta, \phi)$ is a dipole term and is excited by the linear translational motion of the colloidal cluster along the direction $x$. The dipole $p_{\parallel}$ is related to particle velocity $v_{x}$ (linear force $\mathbf{F}$) and can be represented as
\begin{equation}
    p_{\parallel}(t) = \int_{0}^{t} dt^\prime \chi_{T}(t - t^\prime) F(t').
    \label{eq_par_first}
\end{equation}
When the particle is additionally put into rotation by an external torque $\Gamma$, this leads to another density gradient $p_{\perp}$ perpendicular to the direction of the propagation. 
The perpendicular density gradient arises because the particle's rotation displaces more fluid structures from the dense front side of the propagation compared to the less dense wake part (see Fig. 3a in the main text). 

The evolution of $p_{\perp}(t)$ which depends on both $\mathbf{F}(t)$ and $\Gamma(t)$ is formally a second order response, 
\begin{eqnarray}
 p_{\perp}(t) &=&  \int_{-\infty}^{t} \int_{-\infty}^{t^\prime} dt^\prime dt^{\prime \prime} \chi^{(2)}(t , t^\prime, t^{\prime\prime}) \Gamma(t') F(t''). 
 \label{eq_second+order}
\end{eqnarray}
with $\chi^{(2)}(t , t^\prime, t^{\prime\prime})$ the (unknown) second order susceptibility.  We may also assume that $p_{\perp}(t)$ can be written as a first order response, by approximating a  coupled source term with $p_{\parallel}$ and torque $\mathbf{\Gamma}$,
\begin{eqnarray}
 p_{\perp}(t) &\approx &  \int_{-\infty}^{t} dt^\prime \chi_R(t - t^\prime) \Gamma(t') p_{\parallel}(t').
\label{eq_perp_first}
\end{eqnarray}
Combining Eq.~(\ref{eq_second+order}) and Eq.~(\ref{eq_perp_first}), we obtain  
   \begin{eqnarray}
    p_{\perp}(t) &\approx&   \int_{-\infty}^{t}  dt^{\prime \prime} \int_{-\infty}^{t'}  dt^{\prime}  \chi_R(t - t^{\prime})  \chi_T(t^{\prime} - t'') \Gamma(t')  F (t'').
    \label{eq_perp_comb}
\end{eqnarray} 
Comparing Eq.~(\ref{eq_second+order}) and Eq.~(\ref{eq_perp_comb}) it shows that the used approximation amounts in decomposing the second order susceptibility as a product of the first order memory kernels as follows:
\begin{eqnarray}
\chi^{(2)}(t , t^\prime, t^{\prime\prime}) &\approx&  \chi_R(t - t^{\prime})  \chi_T(t^{\prime} - t'').
\end{eqnarray}
It should be noted from the form of the second order susceptibility, $\mathbf{F}$ and $\mathbf{\Gamma}$ are surprisingly not identical regarding their contribution to the perpendicular density dipole $p_{\perp}(t)$ as they enter Eq.~(\ref{eq_perp_comb}) in a cascade manner. The parallel dipole has to exist before it can be rotated to form a perpendicular dipole,

The Magnus velocity can be obtained as $\gamma_0 v_y(t) = c p_{\perp}(t)$ by assuming a bare friction coefficient $\gamma_0$ (see below). Here, $c \mathbf{p}$ is the force due to he dipole $\mathbf{p}$ with an unknown coefficient $c$.
Since, both $\chi_T(t)$ and $\chi_R(t)$ are related to the structural relaxation of the dipole density fields along different directions, the memory kernels $\chi_R(t)$ and $\chi_T(t)$ are expected to have identical relaxation times due to the isotropy of the viscoelastic fluid.

Next, we extrapolate the ideas developed above to obtain the memory kernels and further understand the magnitude of the Magnus effect.

\subsection*{Supplementary Note 2: Extracting memory kernels from experiments}
The experimentally obtained recoil displacement curves of the colloidal cluster provides an amenable way to obtain the memory kernels. Below, we provide a brief calculation.

Consider the experiment with external force $F$ applied to the colloidal cluster along the $x$-direction with no applied torque, $\Gamma = 0$.  At $t = 0$, the force $F$ is switched off and the colloidal cluster is allowed to recoil. The force on the colloid cluster can be written as
\begin{equation*}
  F(t) = \begin{cases}
         F ~~~~~~~~~~~~  t < 0,
        \\
         0 ~~~~~~~~~~~~~ t \geq 0.
        \end{cases}
 \end{equation*}
The dipole $p_{\parallel}$ can be  now written using Eq.~(\ref{eq_par_first}) as
\begin{eqnarray}
\nonumber
p_{\parallel}(t) &=& F \left[ \int_0^{\infty} \chi_T(t') dt' -  \int_0^{t} \chi_T(t') dt'\right] \\
&=& F \left[ \tilde{\chi}_T(s = 0)  -  \int_0^{t} \chi_T(t') dt'\right].
\label{eq_SI_memory_trans_recoil}
\end{eqnarray}
Here, $\tilde{\chi}_T(s = 0)$ is the Laplace transform of the memory kernel $\chi_T(t)$ at $s=0$. Exploiting the fact that $\gamma_0 v_x(t) = c p_{\parallel}$, one obtains the memory kernel $\chi_T$ by double differentiating both sides Eq.~(\ref{eq_SI_memory_trans_recoil}) as follows for $t > 0$
\begin{eqnarray}
    c F \chi_T(t) = -\gamma_0 \frac{d v_x(t)}{dt} =  -\gamma_0\frac{d^2 x(t)}{dt^2}.
\end{eqnarray}
The translational memory kernel can then be obtained from the translational recoil $x(t)$ of the colloidal cluster. 

Next, we consider the steady state of the observed colloidal cluster in presence of the torque $\Gamma$ and the external force $\mathbf{F}$ along $x$ direction. The colloidal cluster displays deflection in the $y$ direction in the steady state. At $t = 0$, the torque is now switched off while keeping $\mathbf{F}$ switched on. The colloidal cluster velocity along $y$ then slows down.

In  experiments, we have  $\mathbf{F}(t) = F$ along $x$-direction and the torque $\Gamma(t)$ can be written as follows 
\begin{equation*}
  \Gamma(t) = \begin{cases}
        \Gamma ~~~~~~~~~~~  t < 0,
        \\
         0 ~~~~~~~~~~~~ t \geq 0.
        \end{cases}
 \end{equation*}
The perpendicular density dipole along the $y$ direction can then be computed via Eq.~(\ref{eq_second+order}) and Eq.~(\ref{eq_perp_first}).
To extract $\chi_R$, we can  then write the expression of Magnus velocity $v_y$ as mentioned below,
\begin{eqnarray}
\nonumber
\gamma_0 v_y(t) &=& c \left({F \Gamma \tilde{\chi}_T (s = 0)}\right) \left[ \int_0^{\infty} \chi_R(t') dt' -  \int_0^{t} \chi_R(t') dt'\right] \\
\nonumber
&=&   c \left({F \Gamma \tilde{\chi}_T (s = 0)}\right) \left[ \tilde{\chi}_R(s = 0)  -  \int_0^{t} \chi_R(t') dt'\right].\\
\end{eqnarray}
Differentiating both sides twice, one obtains the memory kernel from the velocity along $y$,
\begin{eqnarray}
   c \Gamma \left(F \tilde{\chi}_T(s = 0)\right) \chi_R(t) = -\gamma_0\frac{d v_y(t)}{dt} =  -\gamma_0\frac{d^2 y(t)}{dt^2}. 
   \label{eq_extract_rot_kernel}
\end{eqnarray}
The memory kernel $\chi_R(t)$ can hence be obtained from the slowdown of the Magnus (i.e. vertical) displacement after the torque is switched off.

\newpage

\subsection*{Supplementary Note 3: Relation between angle of deflection and dipole angle }

We begin by writing force balance equation of motion for the colloidal cluster in the directions parallel and perpendicular to the external driving force $\mathbf{F}$.  The force due to the presence of dipole is as before, given by $c\mathbf{p}$. As mentioned in the earlier sections, $\gamma_0$ is the bare friction coefficient of the solvent. 
In steady state, i.e., without accelerations, in the described simple picture, we have the following force balances   

\begin{eqnarray}
\nonumber
    -\gamma_0 v_{x} + c p_{\parallel} + F = 0, \\
-\gamma_0 v_{y} + c p_{\perp} = 0. 
\end{eqnarray}
Considering first the velocity parallel to $\mathbf{F}$ in steady driving 
$$ v_{x} =\frac{ c p_{\parallel} + F}{\gamma_0}. $$
When switching off the external force, the recoil velocity parallel to $\mathbf{F}$  immediately after switch off reads
$$ v_{x,o} =\frac{ c p_{\parallel} }{\gamma_0}. $$
The upper two equations allow us to determine the ratio between $p_{\parallel}$ and $F$,
$$ -\frac{v_{x,o}}{v_{x}}\equiv r =-\frac{ c p_{\parallel} }{F+cp_{\parallel}}. $$
One may then obtain the relation between $F$ and the $p_{\parallel}$,
\begin{eqnarray}
    F &=& -cp_{\parallel}\frac{1+r}{r}.
    \label{eq_rel_cp_F}
\end{eqnarray}
Using Eq.~(\ref{eq_rel_cp_F}), we may now estimate the ratio between the velocities  parallel and perpendicular to the force $\mathbf{F}$
\begin{align}
\frac{v_{y}}{v_{x}} = \frac{c p_{\perp}}{ c p_{\parallel}(1 -\frac{1+r}{r})} = -r \frac{p_{\perp}}{p_{\parallel}}.
\end{align}
For small rotation speed $\omega$, $p_\perp/p_\parallel= \theta_p$, and the ratio $v_{y}/v_{x}$ is given by 
\begin{align}
\left\vert\frac{v_{y}}{v_{x}}\right\vert=r \theta_p=rC\omega \tau,
\label{eq_rel_thetap_thetad}
\end{align}
as given in the main text.

For larger values of $\omega$, we may obtain a bound for $v_{y}/v_{x}$ by assuming that $p_\perp$ is in magnitude at most as large as the original $p_\parallel$, i.e., (the first inequality assumes that $v_{x}$ increases in magnitude with $\omega$)
\begin{align}
\left\vert\frac{v_{y}}{v_{x}}\right\vert \lesssim \left\vert\frac{v_{y}(\omega)}{v_{x}(\omega=0)}\right\vert \lesssim r.\label{eq:2}
\end{align}
We may provide a rough estimate for the frequency where saturation sets in by equating  Eqs.~\eqref{eq_rel_thetap_thetad} and \eqref{eq:2}. This yields $\omega_{\rm sat}\approx 1/ C\tau$.

\newpage
\makeatletter
\renewcommand{\figurename}{Supplementary Fig.}
\makeatother
\setcounter{figure}{0}

\begin{figure*}
  \centering
  \includegraphics[width=1\columnwidth]{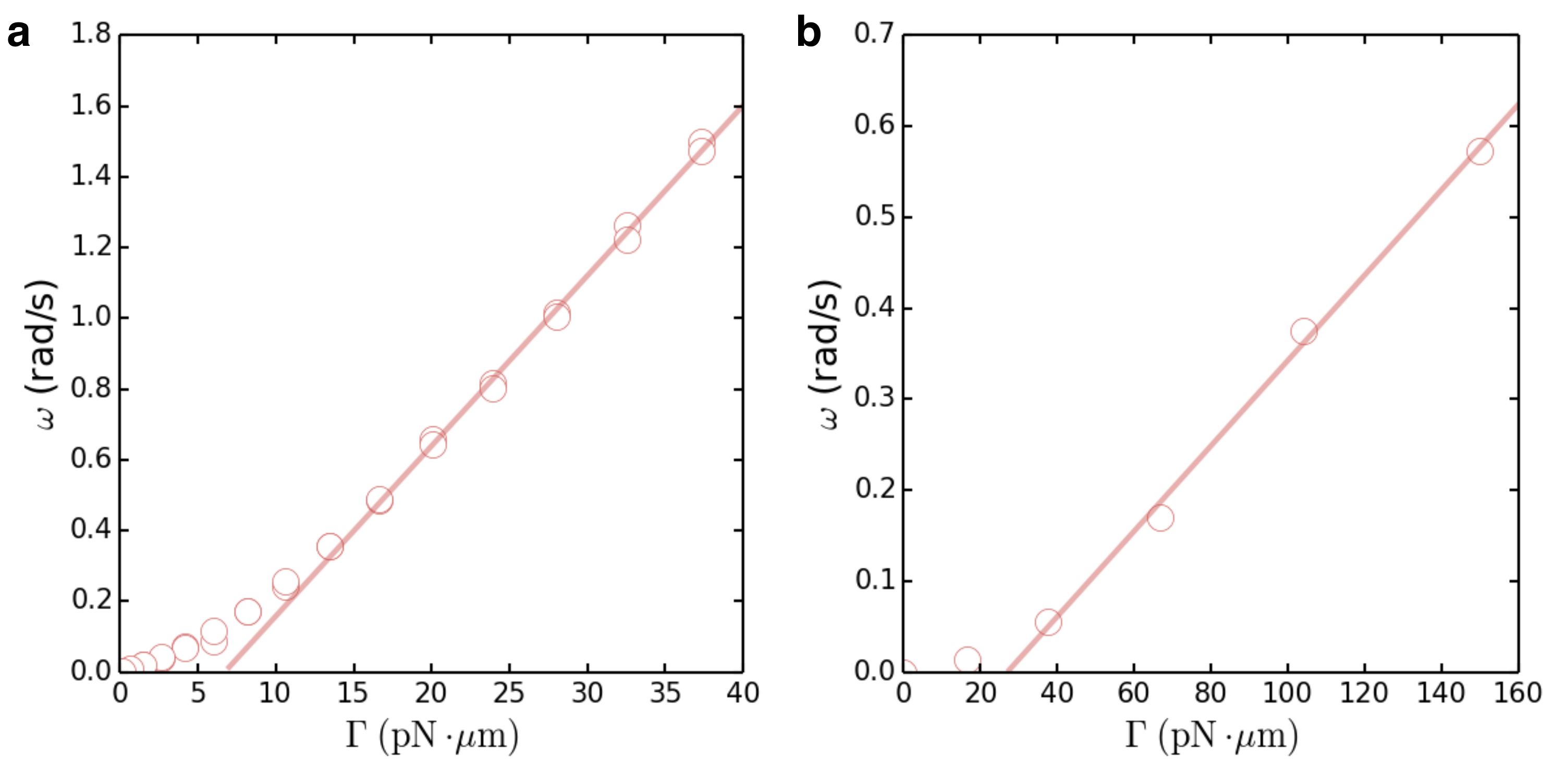}
\caption{\textbf{Calibration of the magnetic torque.} \textbf{a,} The measured rotating speed $\omega$ of a colloid trimer in water as a function of the applied rotating magnetic field strength $H$. The circles are experimental data points and the line is a fit to the equation $\omega=kH^2$ with fitted $k=(3.26\pm0.02)\times10^{-5}~\mathrm{m}^2/(\mathrm{A}^2\cdot\mathrm{s})$. \textbf{b,} The same data as in (\textbf{a}) but now plotted as a function of the applied magnetic torque $\Gamma=\gamma_\mathbf{m}H^2$ as defined in the methods section of the main text. Here $\gamma_\mathbf{m}=6\pi\eta\sigma^3k=(6.99\pm0.02)\times10^{-5}~\mathrm{pN}\cdot\mu\mathrm{m}/(\mathrm{A/m})^2$. 
}
\label{figs1}
\end{figure*}

\begin{figure*}
  \centering
  \includegraphics[width=1\columnwidth]{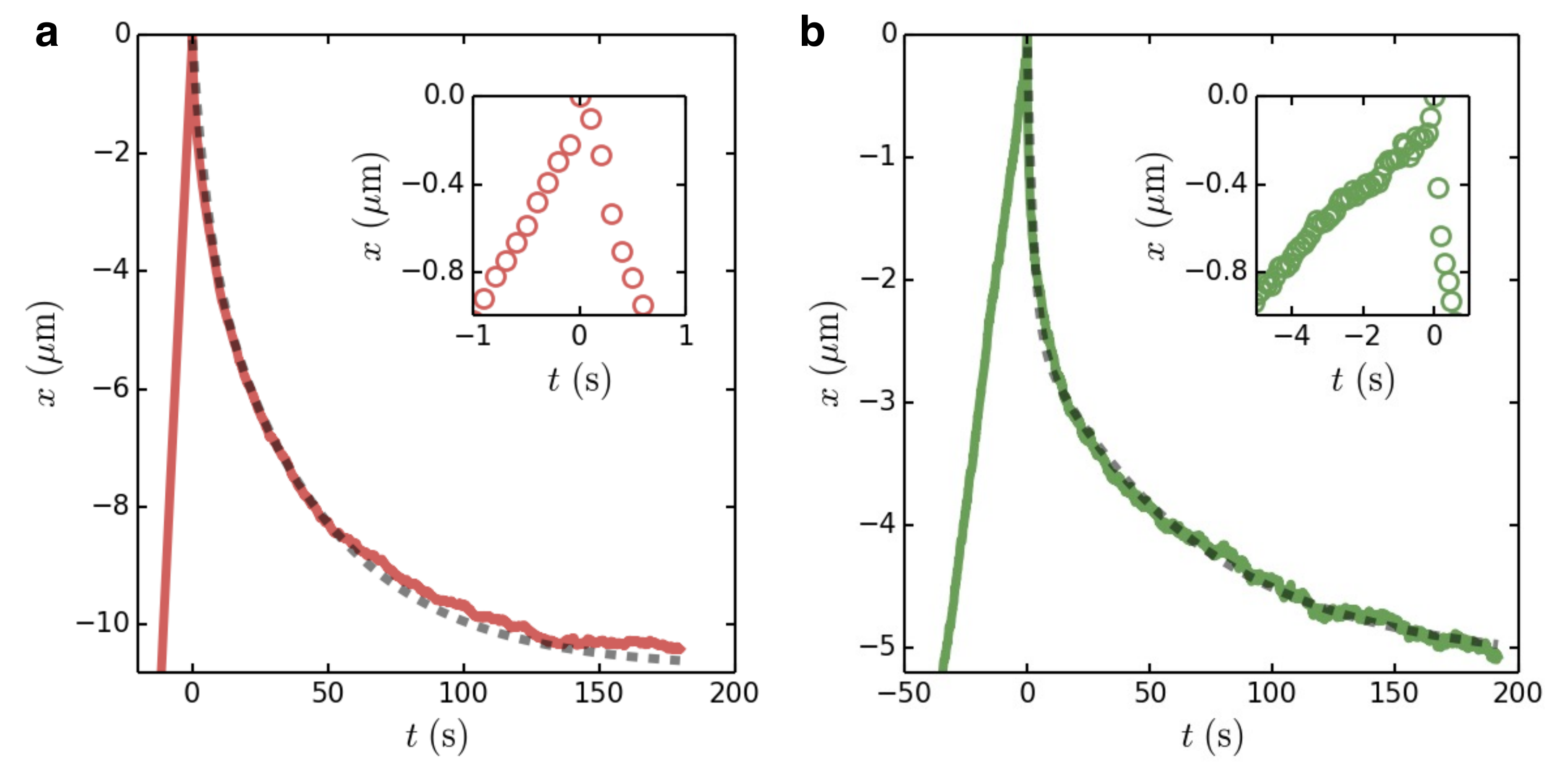}
\caption{\textbf{Characterizing the recoil and relaxation of the viscoelastic fluids.} \textbf{a,} The translational recoil of a colloid trimer in micellar solution, when the applied external force is suddenly removed. The solid lines are experimental data, while the dashed lines are fits to the double exponential function $x=a_1\exp(-t/\tau_1)+a_2\exp(-t/\tau_2)-a_1-a_2$, with fitted $a_1=3.14~\mu\mathrm{m}$, $\tau_1=5.60$ s and $a_2=7.60~\mu\mathrm{m}$, $\tau_2=44.5$ s. Inset: Magnified region near $t=0$. The slope ratio after and before the recoil gives the recoil ratio, which is $r=1.8\pm0.1$. \textbf{b,} The translational recoil of a colloid trimer in polymer solition, when the applied external force is suddenly removed. The solid lines are experimental data, while the dashed lines are fits to the double exponential function $x=a_1\exp(-t/\tau_1)+a_2\exp(-t/\tau_2)-a_1-a_2$, with fitted $a_1=2.56~\mu\mathrm{m}$, $\tau_1=2.56$ s and $a_2=2.74~\mu\mathrm{m}$, $\tau_2=67.5$ s. Inset: Magnified region near $t=0$. The slope ratio after and before the recoil gives the recoil ratio, which is $r=10.5\pm2.3$.
}
\label{figs2}
\end{figure*}

\begin{figure*}
  \centering
  \includegraphics[width=0.8\columnwidth]{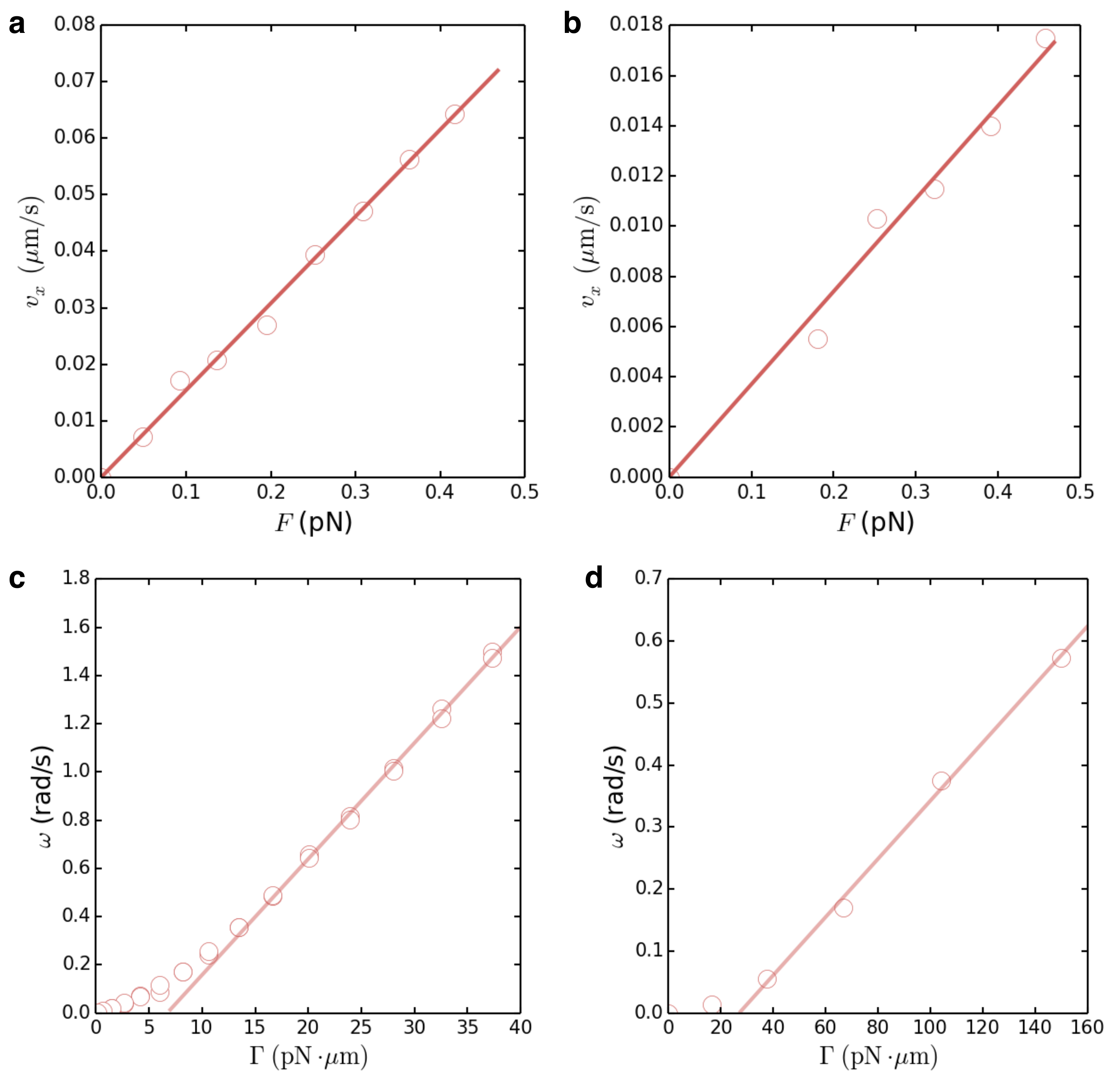}
\caption{\textbf{The translational and rotational friction coefficients.} \textbf{a,} Data points are the measured velocity $v_x$ of a colloid trimer in the micellar solution as a function of the applied force $\mathbf{F}$ in the $x$ direction. The line is a fit to the equation $v_x=F/\gamma$ with fitted $\gamma =6.50\pm0.09~\mathrm{pN}\cdot\mathrm{s}/\mu\mathrm{m}$. \textbf{b,} Data points are the measured velocity $v_x$ of a colloid trimer in the polymer solution as a function of the applied force $\mathbf{F}$ in the $x$ direction. The line is a fit to the equation $v_x=F/\gamma$ with fitted $\gamma =27.0\pm0.8~\mathrm{pN}\cdot\mathrm{s}/\mu\mathrm{m}$.
\textbf{c,} Data points are the measured rotating speed $\omega$ of a colloid trimer in the micellar solution as a function of the applied torque $\Gamma$. A shear thinning behavior is observed around $\Gamma\sim5$ to $15~\mathrm{pN}\cdot\mu\mathrm{m}$. While in our experiments we mainly consider $\Gamma>15~\mathrm{pN}\cdot\mu\mathrm{m}$, where a linear behavior is observed. The solid line is a fit to the equation $\omega=\Gamma/\gamma_\mathrm{R}-\omega_0$ at $\Gamma>20~\mathrm{pN}\cdot\mu\mathrm{m}$, with fitted $\gamma_\mathrm{R}=20.8\pm0.3~\mathrm{pN}\cdot\mu\mathrm{m}\cdot{s}$ and $\omega_0=0.32\pm0.02$ rad/s. 
\textbf{d,} Data points are the measured rotating speed $\omega$ of a colloid trimer in the polymer solution as a function of the applied torque $\Gamma$. A shear thinning behavior is observed around $\Gamma\sim30~\mathrm{pN}\cdot\mu\mathrm{m}$. While in our experiments we mainly consider $\Gamma>30~\mathrm{pN}\cdot\mu\mathrm{m}$ for the polymer fluid, where a linear behavior is observed. The solid line is a fit to the equation $\omega=\Gamma/\gamma_\mathrm{R}-\omega_0$ at $\Gamma>30~\mathrm{pN}\cdot\mu\mathrm{m}$, with fitted $\gamma_\mathrm{R}=213\pm8~\mathrm{pN}\cdot\mu\mathrm{m}\cdot{s}$ and $\omega_0=0.13\pm0.02$ rad/s. 
}
\label{figs3}
\end{figure*}

\clearpage
\noindent

\end{document}